\documentclass[preprint,preprintnumbers,amsmath,amssymb,superscriptaddress]{revtex4}
\usepackage{txfonts}
\usepackage{graphicx}
\usepackage{dcolumn}
\usepackage{bm}
\usepackage{array}
\usepackage[colorlinks=true,plainpages=false,linkcolor=blue,urlcolor=blue,citecolor=blue,pdfpagemode=UseNone,pdfstartview=FitBH]{hyperref}
\begin{document}

\title{Possible Star-of-David pattern charge density wave with additional modulation in the kagome superconductor CsV$_3$Sb$_5$}

\author{J. Luo}
\email{junluo@iphy.ac.cn}
\affiliation{Institute of Physics, Chinese Academy of Sciences,\\
 and Beijing National Laboratory for Condensed Matter Physics,Beijing 100190, China}

\author{Z. Zhao}
\affiliation{Institute of Physics, Chinese Academy of Sciences,\\
 and Beijing National Laboratory for Condensed Matter Physics,Beijing 100190, China}
\affiliation{School of Physical Sciences, University of Chinese Academy of Sciences, Beijing 100190, China}

\author{Y. Z. Zhou}
\affiliation{Institute of Physics, Chinese Academy of Sciences,\\
 and Beijing National Laboratory for Condensed Matter Physics,Beijing 100190, China}
\affiliation{School of Physical Sciences, University of Chinese Academy of Sciences, Beijing 100190, China}

\author{J. Yang}
\affiliation{Institute of Physics, Chinese Academy of Sciences,\\
 and Beijing National Laboratory for Condensed Matter Physics,Beijing 100190, China}

\author{A. F. Fang}
\affiliation{Derpartment of Physics, Beijing Normal University, Beijing 100875, China}

\author{H. T. Yang}
\email{htyang@iphy.ac.cn}
\affiliation{Institute of Physics, Chinese Academy of Sciences,\\
 and Beijing National Laboratory for Condensed Matter Physics,Beijing 100190, China}
\affiliation{School of Physical Sciences, University of Chinese Academy of Sciences, Beijing 100190, China}
\affiliation{Songshan Lake Materials Laboratory, Dongguan, Guangdong 523808, China}

\author{H. J. Gao}
\affiliation{Institute of Physics, Chinese Academy of Sciences,\\
 and Beijing National Laboratory for Condensed Matter Physics,Beijing 100190, China}
\affiliation{School of Physical Sciences, University of Chinese Academy of Sciences, Beijing 100190, China}
\affiliation{Songshan Lake Materials Laboratory, Dongguan, Guangdong 523808, China}
\affiliation{CAS Center for Excellence in Topological Quantum Computation,\\
University of Chinese Academy of Sciences, Beijing 100190, China}

\author{R. Zhou}
\email{rzhou@iphy.ac.cn}
\affiliation{Institute of Physics, Chinese Academy of Sciences,\\
 and Beijing National Laboratory for Condensed Matter Physics,Beijing 100190, China}
\affiliation{Songshan Lake Materials Laboratory, Dongguan, Guangdong 523808, China}

\author{Guo-qing Zheng}
\affiliation{Institute of Physics, Chinese Academy of Sciences,\\
 and Beijing National Laboratory for Condensed Matter Physics,Beijing 100190, China}
\affiliation{Department of Physics, Okayama University, Okayama 700-8530, Japan}

\date{\today}

\begin{abstract}
{$A$V$_3$Sb$_5$ ($A$ = K, Rb, Cs) is a novel kagome superconductor coexisting with the charge density wave (CDW) order. Identifying the structure of the CDW order is crucial for understanding the exotic normal state and superconductivity in this system. Here, we report $^{51}$V nuclear magnetic resonance (NMR) and $^{121/123}$Sb nuclear quadrupole resonance (NQR) studies on kagome-metal CsV$_3$Sb$_5$. Below the CDW transition temperature $T_\textrm{CDW} \sim$ 98 K, an abrupt change of spectra was observed, indicating that the transition is of the first order. By further analyzing the spectra, we find that the CDW order is commensurate. And most remarkably, the obtained experimental results suggest that the charge modulation of the CDW order is of star-of-David pattern and accompanied by an additional charge modulation in bulk below $T^* \sim$ 40 K. Our results revealing the unconventional CDW order provide new insights into $A$V$_3$Sb$_5$.
}
\end{abstract}



\maketitle

\section{Introduction}
Compounds with kagome or honeycomb lattices provide a rich material base for exploring exotic physical phenomena, including topological electronic states\cite{graphene}, highly frustrated magnetisms\cite{Fe3Sn2,Co3Sn2S2,Julien}, and quantum spin liquids \cite{spinliquidReview,spinliquidHerbertsmithite,spinliquiddensitymatrix,Feng2017,Zheng2017}. They are perfect platforms to study the relation between topology, frustration, and electron correlation. However, materials with geometrically frustrated lattices showing superconductivity are still very limited.

Recently, a new transition metal family $A$V$_3$Sb$_5$ ($A$ = K, Rb, Cs) with perfect vanadium kagome-net was found\cite{AV3Sb5}. Angle-resolved photoemission spectroscopy (ARPES), Shubnikov de Haas (SdH) oscillations and density functional theory (DFT) studies categorize $A$V$_3$Sb$_5$ into a Z$_2$ topological class with non-trivial topological bands\cite{Z2topologicalmetal,SdHoscillation}. 
Furthermore, the $A$V$_3$Sb$_5$ system was found to be superconducting with $T_{\rm c}$ range from 0.9 K to 3 K\cite{Z2topologicalmetal,RbV3Sb5}. Above \emph{T}$_{\rm c}$, another phase transition exists at 80 K $\sim$ 100 K revealed by magnetic susceptibility and resistivity measurements\cite{AV3Sb5,RbV3Sb5}.
Scanning tunneling microscopy (STM), optical spectroscopy and ARPES experiments infer that the phase transition around $T$ = 80 K $\sim$ 100 K is a CDW transition\cite{STMtopologicalchargeorder,Opticalspectroscopy1,Opticalspectroscopy2,ultrafast1,ultrafast2,ARPESCDW,ARPESCDW1,Liu2021}.
In the CDW state, hard X-ray scattering and STM measurements observe a 2$a_0$ $\times$ 2$a_0$ $\times$ 2$c_0$ superlattice\cite{STMCDWandZBCP, STMtopologicalchargeorder,Hardxray, PDW,Shumiya2021,WangZW2021}. A star-of-David distortion, found in the known CDW of 1$T$-TaS$_2$\cite{Wilson1975}, was assumed in the kagome plane\cite{STMCDWandZBCP,STMtopologicalchargeorder,Zhao2021}.
As the cleavage surface is $A$ or Sb plane, STM measurements can not directly detect the charge modulation in the vanadium kagome plane. Based on the DFT calculations, an inverse deformation to the star-of-David structure, Tri-hexagonal pattern, was suggested to be the distorted structure\cite{ultrafast1,quantumaoscillation,InverseStarofDavid1}.
However, this has not been clarified by any microscopic probe in the kagome plane yet.
Moreover, a 4$a_0$ charge modulation was suggested to emerge inside the CDW state below $T \sim$ 60 K\cite{Zhao2021,PDW}. The additional charge modulation along the $c$-axis was also reported by X-ray scattering studies\cite{quantumaoscillation}, while not confirmed by other STM and X-ray scattering studies\cite{Hardxray,STMCDWandZBCP}.

Experiments under pressure further elaborated the importance of the CDW order. Transport measurements show that applying pressure can suppress the CDW order and enhance $T_{\rm c}$ by almost three times in $A$V$_3$Sb$_5$\cite{twodome,twodome1,SCReemergence1,SCReemergence2}, making the pressure-temperature phase diagram resemble many unconventional superconductors\cite{cupratephasedigram,Chen2015}. This behavior suggests a close relationship between CDW and superconductivity. The emergence of the pairing density wave in the superconducting state, which might be due to the influence of the CDW order, seems further support this view\cite{PDW}.
In the current stage, the structure of the CDW order in $A$V$_3$Sb$_5$ is still indistinct and identifying it is very important for understanding the origin of the CDW order and its relation to the superconductivity.



In this work, we have performed $^{51}$V nuclear magnetic resonance (NMR) and $^{121/123}$Sb nuclear quadrupole resonance (NQR) measurements on CsV$_3$Sb$_5$. By studying the spectra at low temperatures, we demonstrate that the CDW order is commensurate and our result is consistent with the star-of-David type structural distortion. Below $T^* \sim$ 40 K, a further line splitting is observed in $^{51}$V-NMR spectra, while a line broadening was observed in $^{121}$Sb-NQR spectra. This behavior implies the appearance of an additional charge modulation and further demonstrates the unconventional nature of the CDW order.

\section{Results and Discussions}

\subsection{Charge density wave order: $^{51}$V-NMR results}

\begin{figure}[htbp]
\includegraphics[width=11.8cm]{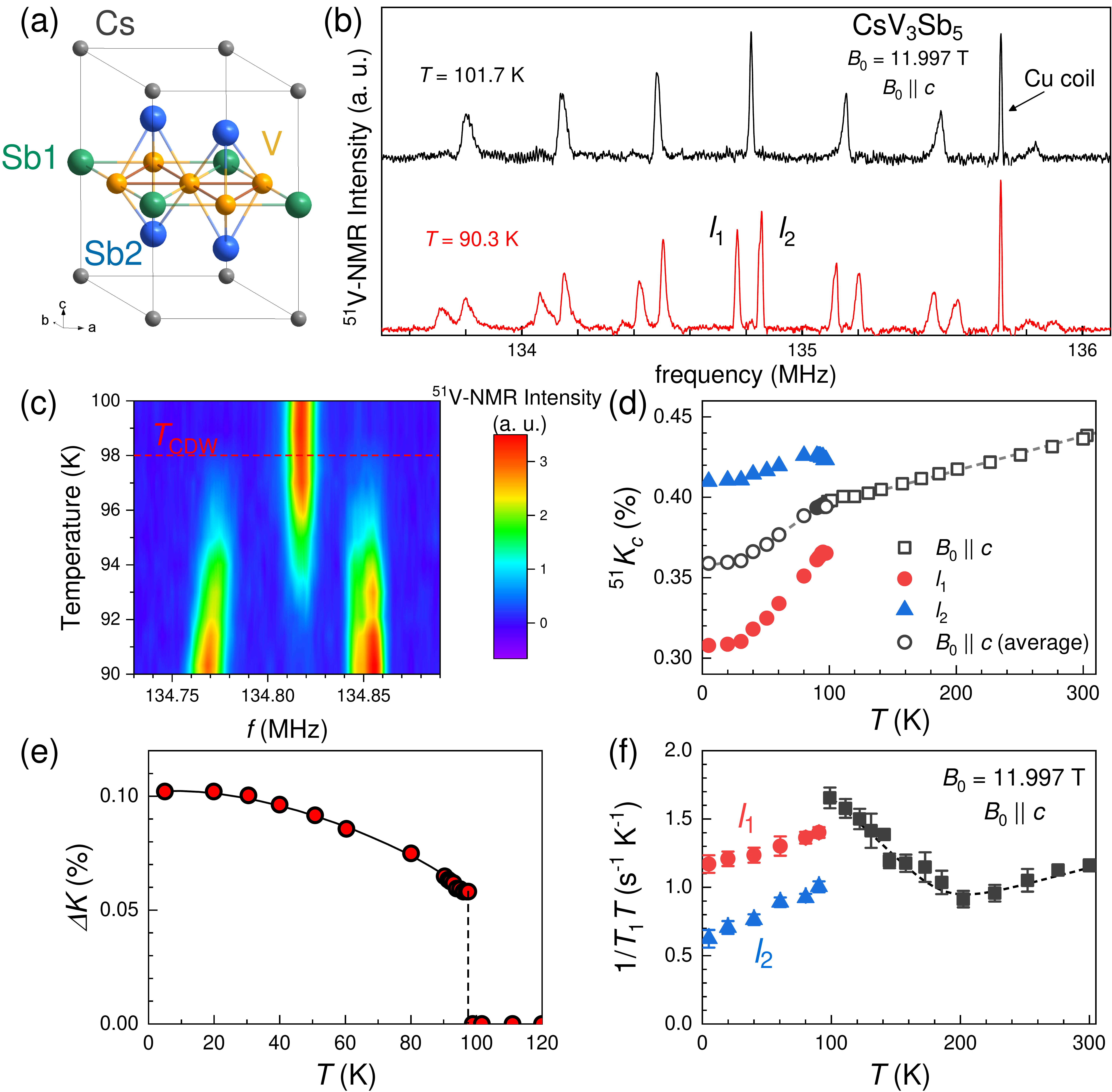}
\centering
\caption{(Color online) \textbf{$^{51}$V-NMR evidence for the CDW order.} (a) The crystal structure of CsV$_3$Sb$_5$ at room temperature. (b) Temperature dependence of the $^{51}$V-NMR spectra above $\textit{T}_\textrm{CDW}$(black color) and below $\textit{T}$$_\textrm{CDW}$ (red color) with the external magnetic field applied along the $c$-axis. The low- and high-frequency central peaks of 90.3 K were assigned as $l_1$ and $l_2$, respectively. The sharp peak around 135.7 MHz is from the Cu coil. (c) Contour plot of the $^{51}$V central lines around $\textit{T}_\textrm{CDW}$, there are three peaks between 98 K and 92 K. (d) Temperature dependence of the Knight shift $K_c$ for $^{51}$V central line.  The average $K_c$ of the two peaks is plotted by black circles. (e) Temperature dependence of the Knight shift difference $\Delta K$ between the two split peaks. (f) Temperature-dependent $1/T_1T$ for $^{51}$V measured at the central lines. Below $\textit{T}_\textrm{CDW}$, $1/T_1T$ and $K_c$ of $l_1$ and $l_2$ are plotted by red dots and blue triangles, respectively. The dash lines are the guides to the eyes. The error bar of $1/T_1T$ is the s. d. in fitting the recovery curve.}
\label{Vspectra}
\end{figure}

In the crystal structure of CsV$_3$Sb$_5$ shown in Fig.~\ref{Vspectra}(a), three V atoms constitute one triangle which is corner-shared to other V triangles, forming a perfect V-kagome lattice. In the unit cell, all V sites are equivalent. A magnetic field of $B_0$ = 11.997 T along the $c$-axis is applied for $^{51}$V-NMR measurements. The total Hamiltonian under magnetic field is\cite{Abragam}:
\begin{equation}
\mathcal{H} = \mathcal{H}_0+\mathcal{H}_Q = -\gamma\hbar \textbf{\textit{I}}\cdot\textbf{\textit{B}}_0(1+K) + \frac{e^2qQ}{4I(2I-1)}[(3I_z^2-I^2)+\frac{1}{2} \eta\left(I_{+}^{2}+I_{-}^{2}\right)]
\end{equation}
where $K$ is the Knight shift, $eq$ = $V_{ZZ}$ = $\frac{ \partial{V^2}}{\partial {Z^2}}$ is the electric field gradient (EFG) along the principal axis, and $V$ is the electric potential, $Q$ is the nuclear quadrupole moment, and $\eta$ = $\lvert$$V_{XX}$ - $V_{YY}$$\lvert$/$\lvert$$V_{ZZ}$$\lvert$ is the asymmetry parameter of EFG. The nuclear quadrupole resonance frequency $\nu_{\rm Q}$ is defined as $\frac{3e^2qQ}{2I(2I-1)h}$. $\gamma$ is the nuclear gyromagnetic ratio, which is 11.193 MHz/T for  the $^{51}$V nucleus.
Since the nuclear spin \textit{I} of $^{51}$V is 7/2, there should appear seven transition lines.
Indeed, the $^{51}$V-NMR spectrum at 101.7 K (above $T_\textrm{CDW}$) shown in Fig.~\ref{Vspectra}(b) consists of one central peak and six satellite peaks.
With decreasing temperature down to 90.3 K, we find that all NMR lines split into two lines with area ratio around 1:1 as illustrated in the bottom of Fig.~\ref{Vspectra}(b). No spectral weight is lost during the transition. We also note that the splitting of the first low-frequency satellite, $\delta f_{satellite}$ = 0.0818 MHz, is different from the splitting of the central peak, $\delta f_{center}$ = 0.0842 MHz, indicating that both magnetic and quadrupole shifts contribute to the observed line splitting. This behavior is the same with the CDW order in YBa$_2$Cu$_3$O$_y$ and Bi$_2$Sr$_{2-x}$La$_x$CuO$_{6+\delta}$ \cite{CDWinYBCO,CDWinBSLCO}, suggesting that a charge modulation occurs at low temperatures.
For the one dimensional (1D) incommensurate CDW order, the NMR spectrum should have two peaks of equal intensities, and most importantly, a continuum between the two peaks\cite{Blinc2002}. For the 2D or 3D cases, if the amplitude of the charge modulation along one direction is much stronger than the others, the NMR spectrum should be similar to the 1D incommensurate CDW order\cite{Blinc2002}. Otherwise, the spectrum should only have one symmetric peak\cite{Blinc2002}.
These features for the incommensurate CDW are not observed in $^{51}$V-NMR spectra, indicating that the observed CDW order is not incommensurate. On the other hand, the commensurate CDW results in discrete peaks\cite{Igor2021}, which is consistent with our observation, proving that the CDW order is commensurate.

To further investigate the CDW evolution, we make a contour plot of the central lines from $T$ = 100 to 90 K, as shown in Fig.~\ref{Vspectra}(c)(the raw data is shown in Supplementary Figure 2).  Three lines are seen between $\textit{T}$ = 98 K and 93 K, implying the coexistence of the CDW and a charge-uniform phase. With decreasing temperature, the NMR intensity gradually shifts from the line of the charge-uniform state to the two split lines related to the CDW state. These are the typical features of the first-order transition. We fit the $^{51}$V-NMR spectra with the Lorentz function to get the resonance frequency $f$ of central peaks. After subtracting the second-order perturbation from the quadrupole interaction (see Supplementary Note 3 for additional analysis), we obtain the temperature-dependent Knight shift $K_{c}$ plotted in Fig.~\ref{Vspectra}(d).
The difference of the Knight shift between $l_1$ and $l_2$ is contributed from the spatial distribution of charge density in the CDW phase. 
$\Delta$$K$ is defined as the Knight shift difference of two split central peaks, and the results are summarized in Fig.~\ref{Vspectra}(e). $\Delta$$K$ abruptly increases to non-zero at 98 K, which is consistent with the temperature dependence of the quadrupole frequency difference $\Delta$$\nu_{\rm Q}$(see Supplementary Figure 4), again suggesting that the CDW transition is of the first order.


We further measure the temperature dependence of $^{51}$V-NMR spin-lattice relaxation rate divided by $T$, $1/T_1T$, as shown in Fig.~\ref{Vspectra}(f). The 1/$T_1T$ decreases with decreasing $T$ down to around 200 K but starts to increase towards $T_{\rm CDW}$.
Since Knight shift decreases with decreasing temperatures, the decrease of $1/T_1T$  at high temperatures might be due to the band effect as Co or Ni-doped BaFe$_2$As$_2$ systems\cite{BaFeCoAs,BaFeNiAs}. At low temperatures, the increase of $1/T_1T$ down to $T_{\rm CDW}$ can be due to either CDW or spin fluctuations.
In view that the CDW transition is of the first order, the contribution from spin fluctuations should not be neglected\cite{NMR2}.
Below $T_\textrm{CDW}$, $1/T_1T$ is measured at both $l_1$ and $l_2$ lines, and both are found to decrease with decreasing temperature due to the gap opening in the CDW state.  
In the end, we note that the averaged Knight shift of two split lines is smaller than the extrapolation from high temperature $K_c$, suggesting a partial gap opening at the Fermi surface in the CDW state.

\begin{figure}[htbp]
\includegraphics[width=11.6 cm]{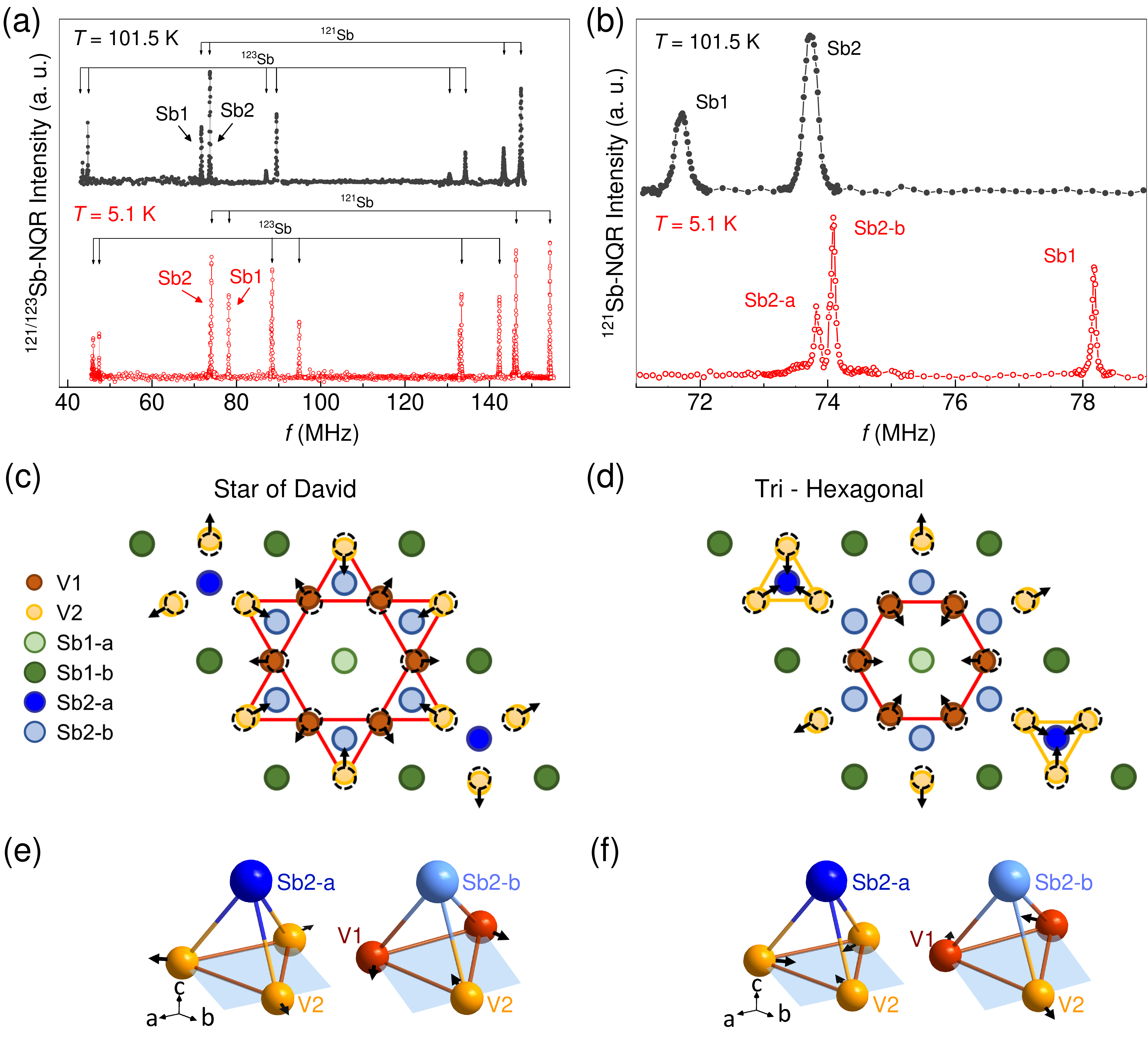}
\centering
\caption{(Color online) \textbf{$^{121/123}$Sb-NQR spectra and illustration of possible CDW patterns.} (a) The NQR spectra of $^{121/123}$Sb at 101.5 K (above $T_\textrm{CDW}$) and 5.1 K (below $T_\textrm{CDW}$). There are five pairs of resonance peaks in the spectrum above $T_\textrm{CDW}$, marked by five pairs of arrows. Each pair of resonance peaks comes from two types of Sb sites in the crystal structure, namely, Sb1 and Sb2. (b) Close-up of the spectrum between 71MHz and 79MHz, corresponding to $\pm$ 1/2 $\leftrightarrow$ $\pm$ 3/2 transitions of $^{121}$Sb. Below the CDW phase transition, one resonance peak from Sb2 splits into two peaks marked as Sb2-a and Sb2-b, and the peak from Sb1 shifts to high frequency. (c) and (d) show the star-of-David and Tri-hexagonal patterns, respectively. The black dashed open circles represent the V site in the pristine crystal structure. The dark yellow and shallow yellow solid circles represent two types of V sites. The olive and shallow olive solid circles represent two types of Sb1 sites. The blue and shallow blue solid circles represent two types of Sb2 sites. (e) and (f) show the local structure around the Sb2 site in star-of-David and Tri-hexagonal patterns, respectively. The black arrows indicate the lattice distortion direction. 
}
\label{Sbspectra}
\end{figure}

\subsection{The structure of CDW order: $^{121/123}$Sb-NQR results}

Next, we use $^{121/123}$Sb-NQR to obtain more information about the structure of the CDW order.  There are two types of Sb sites in CsV$_3$Sb$_5$. Namely, Sb1 is located in the center of the vanadium hexagon, and Sb2 is located above the vanadium triangle, as shown in Fig.~\ref{Vspectra}(a). Sb has two type of isotopes, $^{121}$Sb (\emph{I} = 5/2) and $^{123}$Sb (\emph{I} = 7/2).  For $^{121}$Sb nucleus, the NQR spectrum should have two resonance peaks corresponding to $\pm$ 1/2 $\leftrightarrow$ $\pm$ 3/2 and $\pm$ 3/2 $\leftrightarrow$ $\pm$ 5/2 transitions. For $^{123}$Sb nucleus, the NQR spectrum should have three resonance peaks corresponding to $\pm$ 1/2 $\leftrightarrow$ $\pm$ 3/2, $\pm$ 3/2 $\leftrightarrow$ $\pm$ 5/2 and $\pm$ 5/2 $\leftrightarrow$ $\pm$ 7/2 transitions. So a total of ten lines should be observed in $^{121/123}$Sb-NQR spectrum for CsV$_3$Sb$_5$, which is indeed seen at $T > T_{\rm CDW}$ in the normal state (see Fig.~\ref{Sbspectra}(a)).  
Considering that the atomic ratio between Sb1 and Sb2 is 1 : 4, we assign the lower and higher frequency line in each pair corresponding to Sb1 and  Sb2, as indicated in the figure.
Figure ~\ref{Sbspectra}(b) shows the blow-up to compare the $^{121/123}$Sb-NQR spectrum above and below $T_{\rm CDW}$. With decreasing temperature, an abrupt change of the Sb-NQR spectrum was seen in the CDW state (see Supplementary Figure 5). Unlike a simple splitting observed in the $^{51}$V-NMR spectrum, the two Sb lines evolve into three in the the CDW state. We will show below how to assign these lines, and that the observed change of the Sb-NQR spectrum can be attributed to the star-of-David pattern in the CDW state.

Considering the atomic ratio between Sb1 and Sb2, we can assign that the two peaks around 74 MHz are from the Sb2 site, and another peak is from the Sb1 site. The frequencies of all lines are related to $\nu_{\rm Q}$ and the asymmetry parameter $\eta$. The deduced  $\nu_{\rm Q}$ and  $\eta$ as shown in Table.~\ref{table1} (the detailed calculation is presented in Supplementary Note 6).  As the change of lattice parameter is less than 1\% in the CDW state\cite{AV3Sb5}, the main contribution to the EFG change should be from the unclosed 5$p$ shell of Sb, similar to the case of the O site in the CuO$_2$ plane of the cuprate high-$T_{\rm c}$ superconductors\cite{Zheng1995}. For the Sb1 site, the increase in $\nu_{\rm Q}$ as large as 9\% indicates a strong band renormalization at Sb1-$p$ orbitals in the CDW state. Meanwhile, the absence of a change of $\eta$ suggests that the renormalization should occur at the orbital along the principal axis of EFG, which is along the $c$-axis(see Supplementary Note 7). Therefore, our results suggest a strong band renormalization at  Sb1-$p_z$ orbital in the CDW state, which is consistent with ARPES results\cite{Liu2021}.
In contrast to the Sb1 site, $\nu_{\rm Q}$ of  the Sb2 site changes less than 1\%, suggesting that the band renormalization of the Sb2-$p_z$ orbital is small. However, $\eta$ increases to $\sim$ 0.1 in the CDW state, suggesting a population disparity between Sb2-5$p_x$ and 5$p_y$ orbitals, like the change of the As-4$p$ orbitals inside the nematic state in iron-based high-$T_{\rm c}$ superconductors\cite{Shimojima2010,Zhou2016}. The population disparity between Sb2-5$p_x$ and 5$p_y$ orbitals has not been captured by ARPES measurements so far, probably due to the limited energy resolution of the probe.



\begin{table}[htbp]
\centering
\caption{Experimental results of the quadrupole frequency $\nu_{\rm Q}$ and the asymmetry parameter $\eta$ of the Sb1 and Sb2 sites. The detailed calculation process of $\eta$ can be found in Supplementary Note 6. The unit of $\nu_{\rm Q}$ is MHz. $\eta$ is a dimensionless parameter.}
\label{table1}
~\\
\begin{tabular}{ccccc}
\hline  
$T\ $ & site & $\nu_{\rm Q}\ (^{121}\textrm{Sb})$ &  $\nu_{\rm Q}\ (^{123}\textrm{Sb})$ &  $\eta\ $ \\
\hline
101.5 {\rm K}       & Sb1  &  71.716     &  43.520    &  0  \\
            & Sb2  &  73.728    &  44.763  &  0  \\
\hline
5.1 {\rm K}      & Sb1-a /Sb1-b  & 78.179   & 47.456   & 0 \\
          & Sb2-a & 73.039   & 44.370    & 0.0974 \\
          & Sb2-b & 73.295   & 44.505    & 0.0991 \\

\hline 
\end{tabular}\\
\end{table}

Along with the CDW transition, the star-of-David and Tri-hexagonal types of order as illustrated in Fig.~\ref{Sbspectra}(c) and (d), respectively,  are proposed to be the possible structures\cite{STMtopologicalchargeorder,InverseStarofDavid1,Lin2021}. And based on these two proposed structures, we analyze our experimental results. These two types of order correspond to breathing-in and breathing-out phonon modes of the kagome lattice, respectively. The star-of-David type corresponds to the expansion of Sb1 centering V-hexagon and Sb2 centering V-triangle, and vice versa for the Tri-hexagon type. Both structures form a 2$a_0$ $\times$ 2$a_0$ superlattice. Previous STM and X-ray scattering experiments found that the charge modulation exhibits a $\rm {\pi}$-phase shift along the $c$-axis, suggesting a three-dimensional 2$a_0$ $\times$ 2$a_0$ $\times$ 2$c_0$ charge density wave ordering\cite{STMCDWandZBCP,Hardxray}. For such ordering, the structural distortion pattern within different layers is the same. Moreover, considering that V and Sb1 sit in the V kagome plane and the Sb2 site is close to the plane, the main contribution to the EFG change should be from the in-plane structural distortion. In the analysis below, we only consider the structural distortion in the plane.
In both star-of-David and Tri-hexagonal structures, two V sites exist with the atomic ratio of V1 : V2 site = 1: 1, which explains why $^{51}$V-NMR central peak splits into two peaks with the area ratio of 1 : 1. 
The other feature for the proposed structures is that the Sb2 site should become two distinct sites, namely Sb2-a and Sb2-b as illustrated in Fig.~\ref{Sbspectra} (e) and (f). The atomic ratio of Sb2-a/Sb2-b should be 1 : 3. In our study, we find that the Sb2 peak indeed splits into two peaks with the ratio Sb2-a/Sb2-b around 1 : 3, which is consistent with both star-of-David and Tri-hexagonal patterns. We note that a structure with both star-of-David and Tri-hexagonal characters was reported\cite{quantumaoscillation}. However, the coexistence of the two types of patterns will lead to four distinct Sb2 lines inside the CDW state, which is inconsistent with our observation. We note that more distinct V, Sb1 and Sb2 sites should emerge due to the influence of disorders or CDW domain boundaries. However, this is not observed in our NMR or NQR spectra, as shown in Fig. \ref{Vspectra} and \ref{Sbspectra}. Therefore, our observation suggests that disorders and CDW domain boundaries are rare in our sample.
In principle, the Sb1 site can also become two distinct sites. As the six V atoms are very far from the Sb1 site, however, the difference between Sb1-a and Sb1-b site is very small, making the peak splitting unresolvable in the NQR spectrum.


Next, we distinguish between star-of-David and Tri-hexagonal patterns. For the star-of-David type structure, the average distance between Sb2-a and its nearest V atoms is longer than that between Sb2-b to its nearest V atoms. For the Tri-hexagonal structure, by contrast, the average distance between Sb2-a and its nearest V atoms is shorter than that between Sb2-b to its nearest V atoms. With the smaller distance between the V atom  to the Sb atom, the  influence to the local electron distribution of the Sb site will be more significant.  By EFG calculation, we find that this will enhance the EFG of the Sb2 site(see Supplementary Note 7). 
Our result shows that the $\nu_{\rm Q}$ of Sb2-a is smaller than that of Sb2-b(see Table \ref{table1} and Supplementary Figure 5), suggesting the star-of-David structural distortion in the CDW state.
Our observation is in contrast to the recent reports by combining DFT calculation with ultrafast pump-probe reflectivity experiments\cite{ultrafast1}, STM microscopy images\cite{InverseStarofDavid1}, and quantum oscillation measurements\cite{quantumaoscillation}, in which the Tri-hexagonal type structure distortion was suggested. In these studies, the electron correlation effect is not considered in the DFT calculation. Therefore, the neglected electron-electron interaction could be one possible reason for the contradiction between our and previous studies. However, we note that the displacement of Sb atoms along $c$ axis is not considered in our study based on the star-of-David and Tri-hexagonal patterns. It is also possible that the structure of the CDW pattern is much complex than the current proposed two structures. To further resolve this issue, NMR measurements on Sb sites are needed in the future.


\subsection{Additional charge modulation inside the CDW state}

\begin{figure}[htbp]
\includegraphics[width=13 cm]{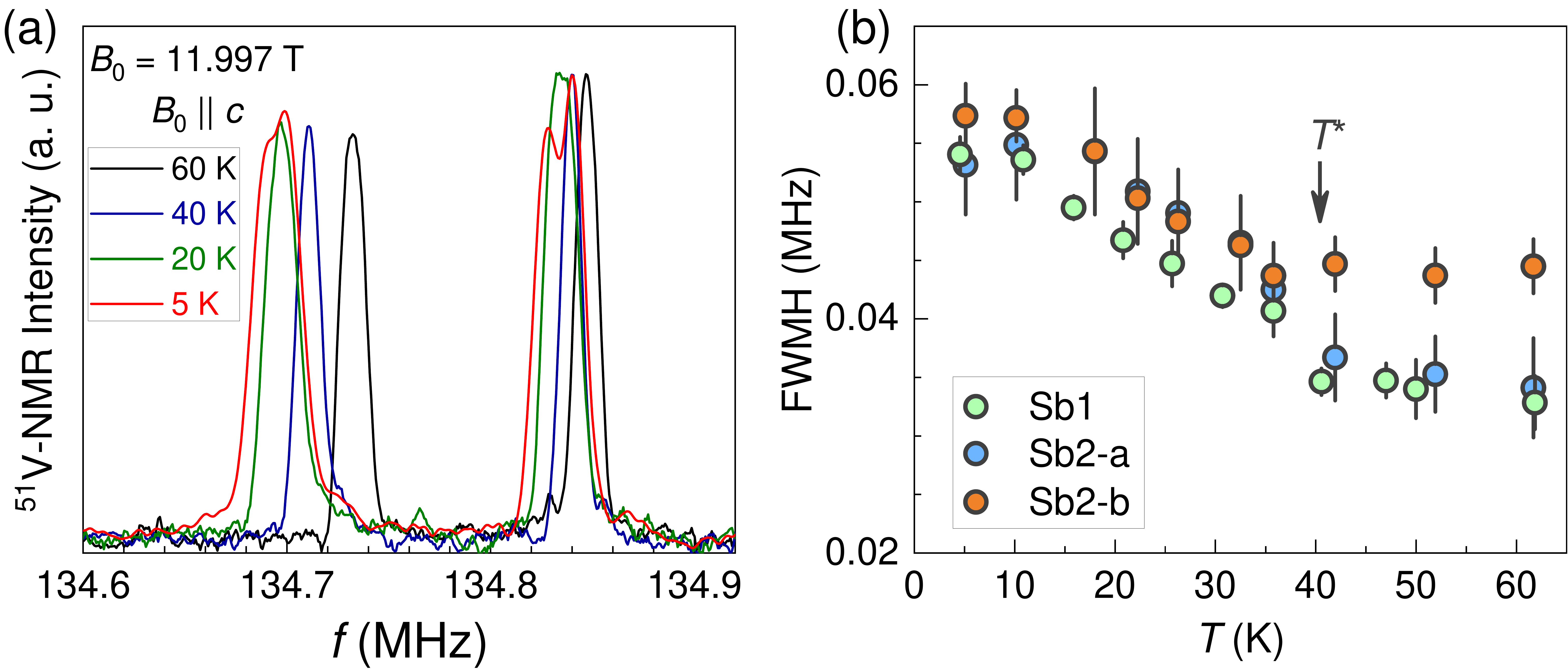}
\centering
\caption{(Color online) \textbf{Evidence for the additional charge modulation inside the CDW state.} (a) Central peaks of $^{51}$V-NMR spectra at various temperatures below $T$ = 60 K. (b) Temperature-dependent full width of half maximum (FWHM) of all Sb sites. The FWHM of both sites start to increase below $T^* \sim$  40 K marked by the black arrow. The error bar is the s.d. in fitting spectra by the Lorentz function.
}
\label{4a}
\end{figure}

Inside the CDW state, we further find another charge modulation at lower temperatures. As shown in Fig.~\ref{4a}(a), two central peaks of the V site start to broaden below $T^*$ $\sim$ 40 K, and further split into four at 5 K. Such further splitting is ten times smaller than the main splitting. So in a less clean sample, only a line broadening can be seen (see Supplementary Figure 10), which also suggests that the splitting is not due to an impurity effect.
The magnetic field is applied along the $c$-axis, which does not introduce any additional symmetry-breaking force.
The possibility of magnetic phase transition can be excluded as there is no anomaly in the temperature dependence of 1/$T_1T$ and the broadening is field-dependent (see Supplementary Figure 11). Thus the change below $T^*$ is mainly due to the Knight shift difference.  

An orbital ordering could also cause a  Knight shift splitting\cite{Zhou2016,FeSe}. Note, however, that the NMR line splitting inside the orbital-ordered state is due to the formation of twinned domains\cite{Zhou2016,FeSe}. In different domains, the applied magnetic field is along different axes, resulting in the line splitting. For a de-twinned sample, only a frequency shift was seen due to the orbital ordering\cite{FeSe2020}. For NQR measurements, on the other hand, no external magnetic field was applied, so the spectra for different domains are the same, resulting in a move of the NQR spectrum below the ordering temperature as seen in LaFeAsO$_{1-x}$F$_x$\cite{Yang}(see Supplementary Note 10). However, we find that the NQR frequencies of all Sb sites increase continuously with decreasing temperature with no anomaly across $T^*$(see Supplementary Figure 5), indicating that our observation is unlikely due to an orbital order. 

The increase of the FWHM of the spectra for all Sb sites below $T^*$, as can be seen in Fig. \ref{4a}(b), suggests that the change is an effect related to local charge density.
We first consider a charge inhomogeneity as a possible cause. An effect due to charge inhomogeneity related to the sample quality would exist already at high temperatures. Given that the line splitting can only be seen at low temperatures, our observation rather implies an emergence of a possible new charge modulation on top of the star-of-David pattern CDW order. The Sb-NQR line width is much broader than the $^{51}$V-NMR linewidth, so only a line broadening is observed down to 5 K. 
Around $T^*$, the temperature evolution of the NMR and NQR spectra is hardly seen. Therefore, we can not confirm whether the additional charge modulation is due to a phase transition or a cross-over. To further investigate this issue, thermodynamic measurements are needed.
Note that the amplitude of this modulation is tiny, which can explain why such additional charge modulation was not observed in some STM and X-ray scattering measurements\cite{Hardxray,STMCDWandZBCP}.

\begin{figure}[htbp]
\includegraphics[width=12 cm]{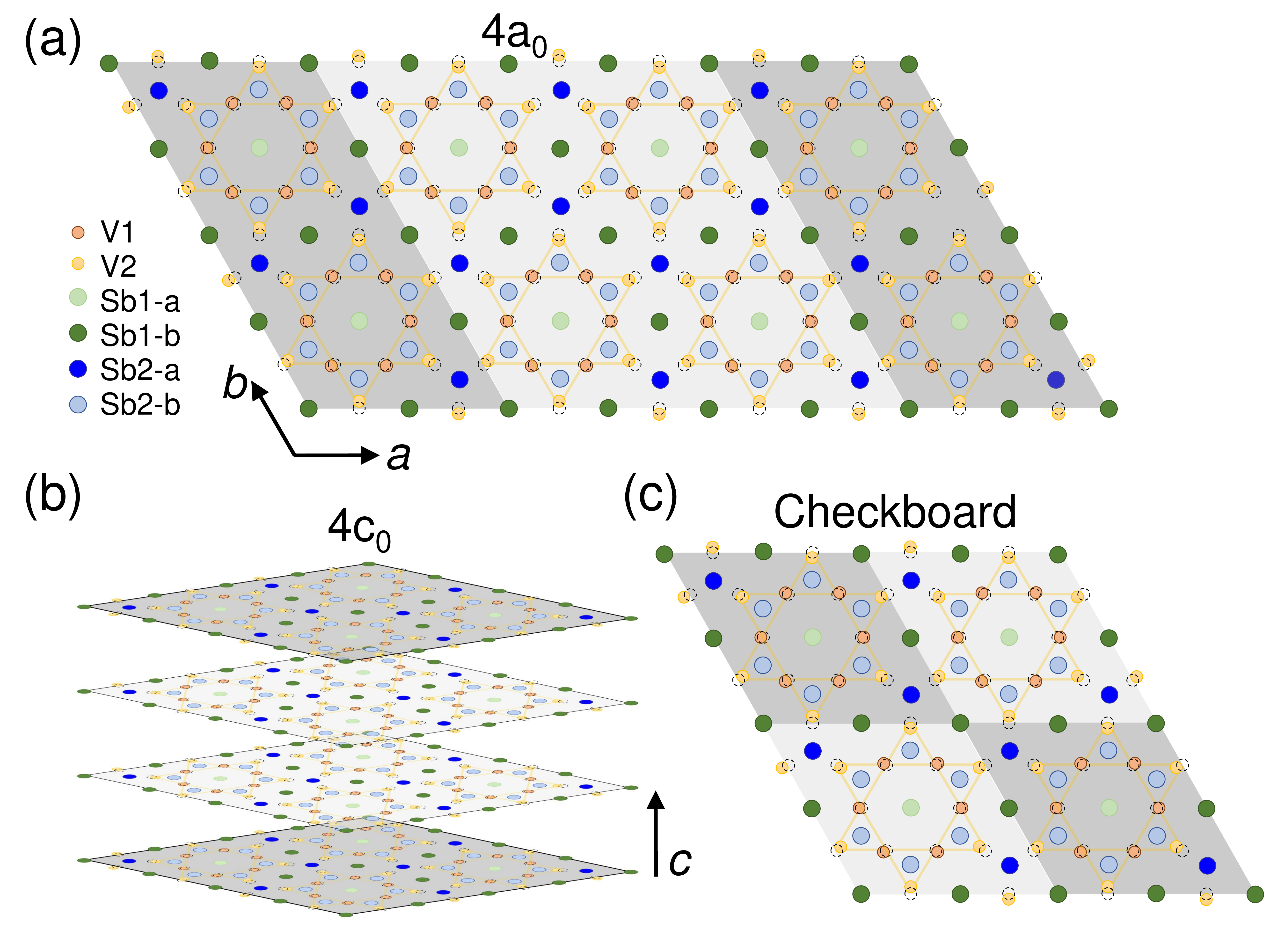}
\centering
\caption{(Color online) \textbf{Illustration of possible CDW patterns below $T^*$.} 4$a_0$, 4$c_0$, and checkboard charge modulations compatible with the NMR spectra under the background of star-of-David type lattice distortion are shown in (a), (b), and (c), respectively. The grey level represents the charge density.
}
\label{Add}
\end{figure}

Several charge modulation patterns that can be compatible with our NMR data, such as the 4$a_0$ \cite{PDW,Zhao2021}, 4$c_0$ \cite{quantumaoscillation}, and checkboard patterns as shown, in Fig. \ref{Add}. 
In this work, we can not distinguish which is the correct pattern. However, we note that the breaking of the in-plane rotational symmetry was observed by measuring $c$-axis resistivity with the in-plane rotation of the magnetic field\cite{Xiang2021}. This suggests that the 4$a_0$ pattern, which breaks the rotational symmetry, is more likely. The 4$a_0$ pattern was also reported by STM measurements\cite{PDW,Zhao2021}. As STM can only obtain the images of surfaces, this 4$a_0$ pattern was suggested to be due to electron correlations related to the surface instability
and electron-phonon interaction\cite{InverseStarofDavid1}. In contrast, NMR detects bulk information, so our results suggest a possible 4$a_0$ pattern in bulk, forming the nematic order\cite{Christensen2021}. Especially, we have noticed that the in-plane anisotropy of the magnetoresistance was observed in the superconducting state, suggesting the  twofold feature of superconductivity\cite{AnisotropicSC,Xiang2021}. Therefore, intertwining between nematicity and superconductivity should be considered for further investigation of the pairing mechanism in CsV$_3$Sb$_5$.

In summary, we have performed NMR and NQR measurements on the kagome superconductor, CsV$_3$Sb$_5$. Below the CDW transition temperature $T_\textrm{CDW}$ = 98 K, the abrupt changes of both $^{51}$V-NMR and $^{121/123}$Sb-NQR spectra indicate that the CDW transition is of the first order. By analyzing the spectra in the CDW state, we suggest that the structural distortion is of the star-of-David type, contrasting with DFT calculations. This implies that electron correlations should be considered for modeling this system. Below $T^* \sim$ 40 K, we further find an additional splitting of $^{51}$V-NMR lines and broadening of $^{121}$Sb-NQR lines, implying the appearance of an additional charge modulation on top of the star-of-David type CDW. All these show that the CDW order in CsV$_3$Sb$_5$ is unique, which will be important in future exploration of the relationship between CDW and superconductivity.

%



~\\
\emph{Note Added: During the preparation of this manuscript, we became aware of two similar NMR works on CsV$_3$Sb$_5$\cite{NMR1,NMR2}. Our $^{51}$V-NMR and $^{121/123}$Sb-NQR spectra are in good agreement with these two works\cite{NMR1,NMR2}. The observation of commensurate CDW order and the first-order CDW transition are consistent with their results. However, knowledge about the charge modulation of the CDW order was  not reported previously. }

\vspace{0.5cm}

\textbf{Methods}

\textbf{Samples}

Single crystal CsV$_3$Sb$_5$ was synthesized by the self-flux method\cite{AV3Sb5}. The typical size of the single crystal is around 3 mm$\times$2 mm$\times$0.1 mm. $T_{\rm c}$ was determined by DC susceptibility measured by a superconducting quantum interference device with the applied field 1 Oe parallel to the $c$-axis (see Supplementary Figure 1). $T_{\rm c}$ is close to 3.5 K, which is among the highest values for this compound, indicating its high quality.

\textbf{NMR and NQR measurements}

$^{51}$V-NMR experiments were performed on one single crystal sample at a fixed magnetic field along the $c$-axis. The spectra were obtained by adding Fourier transforms of the spin-echo signal recorded for regularly spaced frequency values. $^{121/123}$Sb-NQR spectra were measured on a collection of $\sim$ 50 single crystals by sweeping the frequency point by point and integrating spin-echo intensity. By performing EFG calculation, we find that the EFG principal axis of $^{121/123}$Sb is along the $c$-axis(details about EFG calculation is present in Supplementary Note 7). Thus, we arrange the CsV$_3$Sb$_5$ single-crystal flakes along the $c$ direction, ensuring the radio-frequency field $H_1$ in the $ab$ plane(see Supplementary Figure 8).

\vspace{0.5cm}
\textbf{Data Availability}

The data that support the findings of this study are available from the corresponding authors upon reasonable request.

\begin{acknowledgments}
The authors thank Kun Jiang and Hui Chen for valuable discussions. This work was supported by the National Natural Science Foundation of China (Grant Nos. 11974405, Nos. 61888102, Nos. 51771224 and Nos. 11904023), the National Key Research and Development Projects of China (Grant Nos. 2017YFA0302904 , Nos. 2018YFA0305800 and 2019YFA0308500) and the Strategic Priority Research Program of the Chinese Academy of Sciences (Grant Nos. XDB33010100 and Nos. XDB33030100). This work was supported by the Synergic Extreme Condition User Facility.
\end{acknowledgments}


\vspace{0.5cm}
\textbf{Author contributions}

J. Luo and Z. Zhao contributed equally to this work. The single crystals were grown by Z.Z., H.T.Y. and H.J.G. The NMR measurements were performed by J.L., Z.Y.Z., J.Y., A.F.F. and R.Z. R.Z. and G.-q.Z. wrote the manuscript with inputs from J.L. All authors have discussed the results and the interpretation.

\vspace{0.5cm}

\textbf{Competing Interests}

The authors declare no competing interests.

\end{document}